# Black Arsenic as an Optimum Gas Sensor: vdW Corrected Density Functional Theory Calculations

Rameshwar L. Kumawat[†], Milan Kumar Jena[#], Biswarup Pathak*[,†,#,]

[†]Discipline of Metallurgy Engineering and Materials Science, [#]Discipline of Chemistry, School of Basic Sciences, Indian Institute of Technology (IIT) Indore, Indore, Madhya Pradesh, 453552, India

*E-mail: biswarup@iiti.ac.in

**Abstract**

Recent experiments demonstrate the synthesis of 2D black arsenic exhibits excellent electronic and transport properties for nano-scale device applications. Herein, we study by first-principle calculations density functional theory together with non-equilibrium Green's function methods, the structural, electronic, adsorption strength (energy), charge transfer, and transport properties of five gas molecules (CO, $CO_2$, NO, $NO_2$, and $NH_3$) on a monolayer of black arsenic. Our findings suggest optimum sensing performance of black arsenic that can even surpass that of other 2D material such as graphene. Further, we note the optimum adsorption sites for all the five gas molecules on the black arsenic and significant charge transfer between the gas molecules and black arsenic are responsible for optimum adsorption strength. Particularly, the significant charger transfer is a sign that the interaction between the target gas molecule and nano-scale device is sufficient to yield noticeable changes in the electronic transport properties. As a proof of principle, we have examined the sensitivity of a modeled nano-scale device towards CO, $CO_2$, NO, $NO_2$, and $NH_3$ gas molecules, indicating that it is indeed possible to reliably detect all the five gas molecules. Thus, based on all these findings, such as sensitivity and selectivity to all the five gas molecules adsorption make black arsenic a promising material as an optimum gas sensor nano-scale device.

## 1. Introduction

Over decades environmental pollution due to the emission of various toxic, hazardous and greenhouse gases/vapors is an emerging threat to both human health and environmental safety. It has been reported that prolonged exposure to very low concentration (for example $NO_2$ <3 ppm) can cause lethal effects on human health.[1-5] Therefore, sensing of toxic gases/vapors plays a significant role in environmental monitoring, industrial, agricultural, and medical applications.[5] Thus, designing of gas sensing nano-scale devices are becoming increasingly mandatory for such applications. In this respect, semiconducting metal oxide based gas sensors have been broadly used and investigated due to their easy fabrication, good sensitivity towards gas molecules, inexpensive, and energetic consumption.[5-11] However, low stability, poor selectivity, high-temperature operation, and low recovery time arises as to the key restraining aspects for concrete use of such material in sensing applications.[9-11] Because of those disadvantages and inspired by the increasing demands for high sensitivity, selectivity, inexpensive, and stable gas sensors, several new low-dimensional nanomaterials have been continually investigated.[12-25]

Since the inception of graphene,[26,27] much attention has been dedicated to discovering new low-dimensional (2D and 1D) nanomaterials due to their outstanding properties for example high carrier mobility, mechanical strength, bandgap tunability, high-surface-to-volume-ratio, etc.[12-27] Graphene which started as the most popular material among the scientific community has lost some fame due to its often-mentioned lack of bandgap, which limits its applications in the nano-scale devices. Succeeding this prototypical route of graphene,[26,27] numerous other 2D nanomaterials came into practical realization. New 2D materials with a bandgap include silicene,[28,29] stanene,[30] germanene,[31] phosphorene,[32] arsenene,[33] etc. Besides, transition metal

dichalcogenides (TMDCs)[34-36] and MXenes also represent another developing class of 2D nanomaterials.[37-40]

Recently, single layer structures of group V block element (or pnictogens) analogues of graphene have studied tremendously due to their outstanding structural, chemical, physical, electronic and transport properties. Group V elements (N, P, As, Sb, Bi) have been the theme of active research both theoretical and experimental.[33,41-46] In this respect, arsenic has attracted considerable attention due its outstanding structural, chemical and physical properties. The arsenic-based counterpart of black phosphorus (b-P) is so-called black arsenic (b-As). 2D b-As is a metastable form of arsenic with an orthorhombic structure similar to b-P, and its monolayer is called arsenene.[33,45,46] 2D b-As exhibits a puckered honeycomb-like structural configuration with high stability more than black phosphorus.[33] Nikolas and co-workers have demonstrated the synthesis of orthorhombic b-As based on the crystallization of amorphous arsenic by mercury vapors.[47] The monolayer of b-As has a direct bandgap calculated with GGA functional.[33,45,46] It possesses high anisotropic carrier mobility which is comparable with the monolayer of b-P[49] and $MoS_2$.[50] In general, b-As holds great potential for futuristic nano-scale device applications due on their outstanding carrier mobility in monolayer, strong mechanical, and high surface to volume ratio.[33,45-48,51] Keeping all these properties in mind researchers have already been explored several applications, for example, photocatalysis,[52] field-effect transistors,[53] batteries,[54] etc. Sun and co-workers have also demonstrated that puckered arsenene is a very promising material for thermoelectric cooling applications.[48] Similarly, Cuniberti and co-workers have studied the electronic, phonon band structure and transport properties.[55] These findings have paved the way for its application in next-generation nano-scale devices. To the best of our knowledge, gas sensing application of b-As monolayer still remain elusive. Motivated by all these reports, herein, we have conceived the idea that such a puckered structure of b-As monolayer can be a suitable nano-scale device for gas sensing applications.

Further, it would be very interesting to understand whether the puckered nature in b-As is helpful for gas molecule sensing or not. It is therefore essential to realize the interactive nature between the b-As monolayer and adsorbate gas molecules for the possible application of b-As based nano-scale sensing device. To be an efficient sensor, it must have three features: (I) a small but not negligible adsorption energies (i.e., binding energy), (II) sensitivity, and (III) selectivity to different gas molecules. Herein, we study five gases (CO, $CO_2$, NO, $NO_2$, $NH_3$) on the b-As monolayer. Using density functional theory (DFT), we study the structural, electronic, adsorption, and charge transfer properties of the pristine b-As and b-As+gas molecules. Further, we compute the transmission functions and current-voltage signals for b-As+gas molecules using the non-equilibrium Green's function (NEGF) methods in combination with DFT.

## 2. Computational Details

The structural optimization and electronic calculations are done by first-principles DFT calculations using the VASP (Vienna Ab Initio Simulation) code[56] within the GGA-PBE functional.[57] We have used the van der Waals (vdW) corrections employing the Grimme (DFT-D3) method.[58] The vdW corrections are used to delight the interaction between the target gas molecules and the b-As surface. The monolayer of b-As is periodic in the $xy$ plane. We have used 16 Å of vacuum along the $z-$ direction to avoid the interactions between repeated images of the systems. An $11 \times 11 \times 1$ k-grid mesh samples the Brillouin zone integration within the Monkhrast-pack scheme[59] for a $3 \times 3$ supercell. We have chosen an energy cut-off of 400 eV for the plane-wave basis. Projected augmented wave (PAW) potentials are employed to define the ion-electron interactions.[60] All the systems are fully optimized to obtain the energetically most state-structure with residual forces on each atom is smaller than 0.01 eV/Å. The adsorption energies ($E_{ads}$) of the gas molecules is calculated by employing the following equation:

$$E_{ads} = [E_{b-As+gas} - (E_{b-As} + E_{gas})] \tag{1}$$

where $E_{b-As+gas}$ represents the total optimized energy of the b-As+gas system, $E_{b-As}$ and $E_{gas}$ are the energy of isolated b-As surface and the isolated gas, respectively within the geometry of the b-As+gas system.

The electronic charge-density-difference ($\Delta\rho(r)$) is shown to understand the interaction nature of the different gases on the b-As substrate and this is calculated using the below given equation:

$$\Delta\rho(r) = [\rho_{b-As+gas}(r) - \rho_{b-As}(r) - \rho_{gas}(r)] \tag{2}$$

where $\rho_{b-As+gas}(r)$ represents the total charge density on the b-As substrate in the presence of an adsorbed gas, $\rho_{b-As}(r)$ is the charge density on the isolated b-As substrate, and $\rho_{gas}(r)$ is the charge density of the isolated gas. The charge transfer analysis due to the adsorption of all the five gases on the b-As substrate is done using the Bader charge methods as applied in the VASP code.[61] The electronic transport calculations are done using NEGF combined with DFT methodology in the TranSIESTA module of the SIESTA code.[62,63] We have done electronic transport calculations to calculate the sensing capability of pristine b-As monolayer as nano-scale device to detect all the five gas molecules [CO, $CO_2$, NO, $NO_2$, and $NH_3$]. For this purpose, we again fully optimized the pristine b-As and b-As+gas devices in the SIESTA code. The van der Waals interactions (Grimme) have been considered for all calculations. An energy cut-off of 300 Ry and double-ς polarized basis sets have been used and $1 \times 11 \times 56$ k-points grid have been considered during the electronic transport calculations. We have tested the effect of spin polarization for one system (b-As+NO, because NO is paramagnetic gas) but not considered for other gas molecules. For the transport calculations, our proposed nano-scale device is divided into three parts: left (L) and right (R) electrodes and a central scattering region (device region). The electronic transmission function which defines the probability of an incoming electron transferred from the one electrode to the other with the specific energy (E) is calculated as following the given equation

$$T(E,V) = tr[\Gamma_L(E,V)G_C(E,V)\Gamma_R(E,V)G_C^\dagger(E,V)] \qquad (3)$$

where $\mathcal{G}$ is the Green's function of the central region and $\Gamma_{L/R}$ is the coupling matrix of left (L) and right (R) electrodes.[64,65,66]

## 3. Results and Discussion

Firstly, we have optimized the unit cell of the b-As and obtained the lattice constants a = 4.71 Å, and b = 3.69 Å and the As-As bond lengths are $d_1$ = 2.49 Å, and $d_2$ = 2.50 Å in monolayer structure. Our obtained lattice constants and bond lengths are agreed well with the previous results.[46] We have modeled the b-As in a 3 × 3 supercell (**Figure 1**), which has lattice parameters of 11.08 × 14.08 Å. The boundary structure of b-As is anisotropic along the $x$ and $y$ directions, which are marked as zigzag and an armchair directions.[55] The calculated electronic density of states is displayed in **Figure 1**. It is interesting to note that b-As has a bandgap of ~0.67 eV. Our observations are nearly in agreement with previous results.[46] We have observed from the PDOS of b-As that the states near the Fermi level $(E - E_f)$ have contributions from both S and P-type orbitals. However, the maximum contributions come from the P-type orbitals to the total DOS than that from the S-type orbitals.

The adsorption behavior of all the five gas molecules [CO, $CO_2$ NO, $NO_2$, and $NH_3$] on the b-As surface, is examined. For each adsorption system, a gas molecule is placed onto the b-As surface, and the whole (b-As+gas) system is then optimized. We have systematically investigated the adsorption behavior of all the five gases at three possible adsorption sites (top, hollow, and bridge) on the b-As surface. The computed relative energies are tabulated in [**Table S1 Supporting Information** (**SI**)]. The most stable adsorption configurations of b-As+gas systems are shown in **Figure 2(a-e)** and the obtained adsorption heights ($h$) are tabulated in **Table 1**.

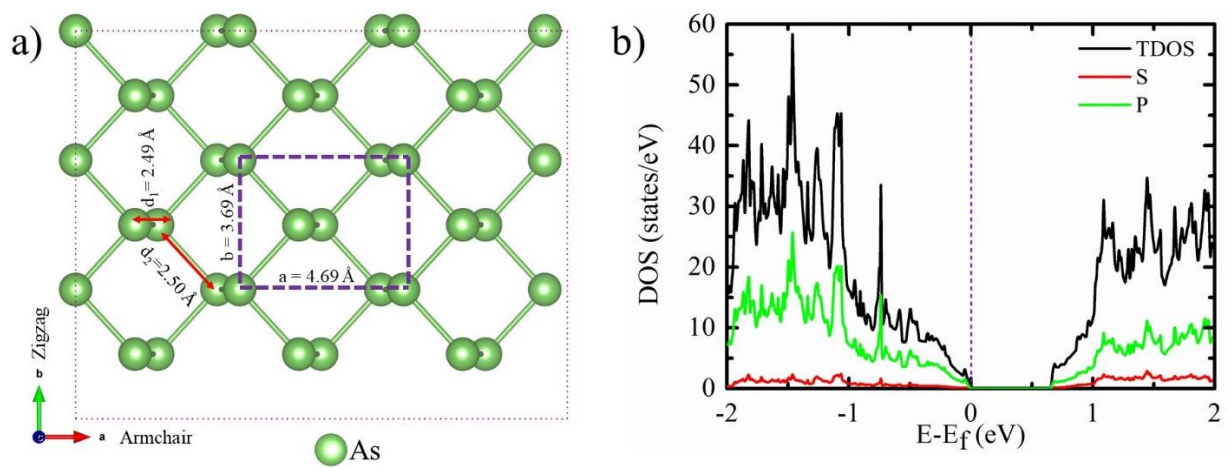

**Figure 1**. (a) Optimized geometric structure and (b) electronic density of states of the b-As monolayer. The Fermi-level is aligned to zero.

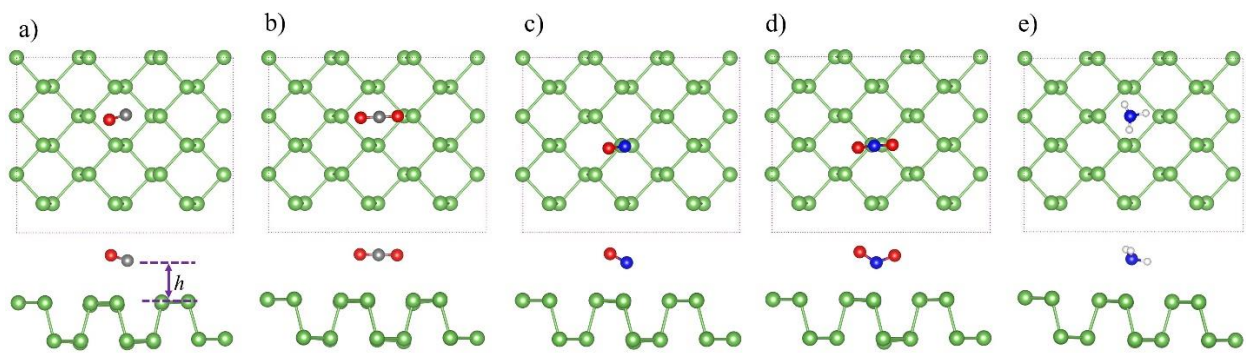

**Figure 2.** Top and side views of the energetically most stable configurations (a) b-As+CO, (b) b-As+ $CO_2$, (c) b-As+NO, (d) b-As+$NO_2$, and (e) b-As+$NH_3$ adsorption. The adsorption height ($h$) between the gas molecule and the b-As surface is shown in (a). The violet dashed rectangle is the supercell in the current work. Atoms color code: As (green), O (red), N (blue), C (grey), and H (light white), respectively.

**Table 1.** Adsorption height ($h$), adsorption energies ($E_{ads}$), and net (Bader) charges on the gas molecules for different b-As+gas systems.

| Gas Molecules | $h$(Å) | $E_{ads}$ (eV) | $\Delta Q$ (|e|) |
|---|---|---|---|
| CO | 2.79 | -0.11 | 0.04 |
| $CO_2$ | 3.01 | -0.13 | 0.04 |
| NO | 2.52 | -0.22 | 0.19 |
| $NO_2$ | 2.43 | -0.82 | 0.46 |
| $NH_3$ | 2.77 | -0.21 | 0.10 |

We have analyzed the binding character of gas molecules with b-As surface systematically. In the case of CO and $CO_2$ molecules, the C atom of CO and $CO_2$ is found at the center of the puckered surface (**Figure 2a, b**). The adsorption height between the b-As surface and the CO molecules (2.79 Å) is smaller than $CO_2$ adsorption (3.01 Å). For dipolar gas molecule NO, it locates to the top site of an As atom (**Figure 2c**). It is noted that the adsorption height of NO molecule with the b-As surface is noticeably reduced to 2.52 Å. In the case of $NO_2$, the adsorption height between the $NO_2$ and the b-As surface is 2.43 Å (**Figure 2d**). In the case of tetratomic molecule $NH_3$, it is also residing above the b-As surface with an adsorption height of 2.77 Å with the N atoms which is sited at the center of the puckered surface (**Figure 2e**).

For a qualitative explanation of the adsorption strength on the b-As surface, we have calculated the adsorption energy ($E_{ads}$) as discussed in the computational method section. The obtained adsorption energy values for all the five gas molecules along with the adsorption height from the b-As surface have been tabulated in **Table 1**. We note that the CO gas molecule has the smallest $E_{ads}$ of -0.11 eV/unit cell, while $NO_2$ molecule has the largest $E_{ads}$ value of -0.82 eV/unit cell among all the five gases investigated in this work (**Table 1**). The $E_{ads}$ values for $CO_2$, $NH_3$, and

NO molecules are -0.13, -0.21, and -0.22 eV/unit cell, are lying in between. Moreover, all these calculated $E_{ads}$ for the b-As surface are sufficient to be affected by the thermal disturbance at room temperature which is at the energy scale of $k_B T$ ($k_B$: Boltzmann constant).[67,73] This shows that the b-As surface is promising for all the five gas molecules detection at room temperature. The nitrogen (N) based gas molecules usually have larger $E_{ads}$ values than those for CO and $CO_2$ $E_{ads}$ values, demonstrating that b-As substrate is relatively more sensitive to the N-based toxic gas molecules than carbon-based, which is alike to black phosphorene,[73] $MoS_2$,[23,70] borophene,[67] germanene,[72] and silicene[71] that are tremendously sensitive to adsorption of $NO_x$ ($x = 1,2$) and $NH_3$. Moreover, we have noted that most of the gas molecules revealed optimum adsorption energy values to the b-As surface compared to other previously reported monolayer materials. We have tabulated the $E_{ads}$ values of all the five gas molecules on different monolayer substrates as collected from the literature along with the b-As in **Table 2**. Seeing the outcomes above, we then go to use these nano-scale devices for gas detection. We find that graphene displays weaker adsorption energy values for all the five gas molecules in the range of meVs[68,69] though this principal to easy recovery, it typically decreases the signal-to-noise ratio.[17] Thus, b-As may be a more suitable gas sensor than graphene because it shows optimum adsorption energy values.

**Table 2**: Comparative summary of adsorption of gases ($E_{ads}$ in eV) on different 2D surfaces collected from literature against the b-As surface.

| Surface \| Gas | CO | $CO_2$ | NO | $NO_2$ | $NH_3$ |
|---|---|---|---|---|---|
| Black Arsenic (b-As) | -0.11 | -0.13 | -0.22 | -0.82 | -0.21 |
| Black Phosphorene[73] | -0.32 | -0.41 | -0.86 | -0.60 | -0.50 |
| Graphene[68,69] | -0.01 | -0.05 | -0.03 | -0.07 | -0.03 |
| $MoS_2$[70,23] | -0.44 | -0.33 | -0.55 | -0.14 | -0.16 |
| Germanene[72] | -0.16 | -0.10 | -0.51 | -1.08 | -0.44 |
| Silicene[71] | -0.18 | -0.04 | -0.35 | -1.37 | -0.60 |
| Borophene[67] | -1.38 | -0.36 | -1.79 | -2.32 | -1.75 |

Furthermore, the adsorption energies reported for CO, CO$_2$, NO, and NH$_3$ gases on the black phosphorene substrate are slightly larger with -0.32, -0.41, -0.86, and -0.50 eV/unit cell, respectively.[73] Similarly, the adsorption energy values reported for the different gas molecules on the MoS$_2$ substrate are somewhat larger for CO, CO$_2$, and NO (-0.44, -0.33, and -0.55 eV/unit cell), respectively.[23,70] Similar qualitative tendency of gas molecule adsorption is also perceived for 2D germanene,[72] silicene,[71] borophene.[67] Hence, form the ongoing discussion, we find that our obtained adsorption energy values are optimum for most of the systems and show enough sensitive nature of b-As towards favourable gas sensing. Therefore, b-As could be a promising candidate for gas sensor applications.

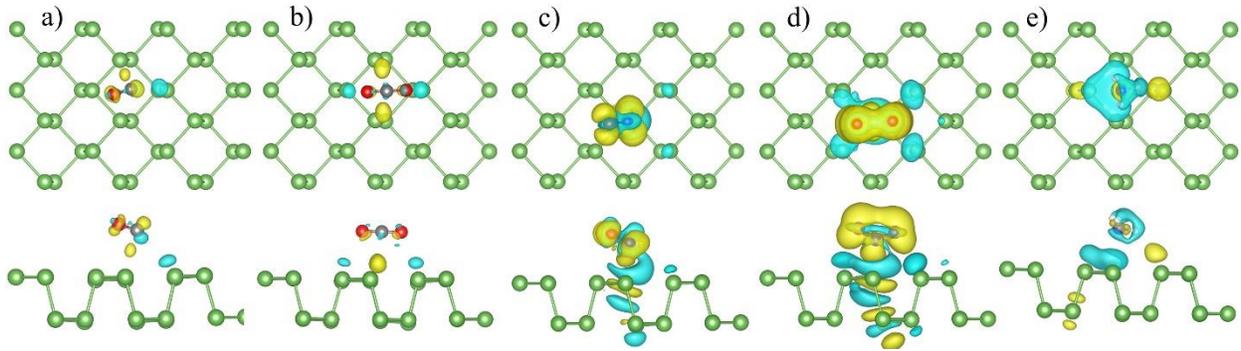

**Figure 3**. (a-e) The adsorption configurations (b-As+gas molecule) and charge transfer for CO, CO$_2$, NO, NO$_2$, and NH$_3$ are plotted. The iso-surface value for all the systems is 0.0006 e/Å$^3$. The blue iso-surface represents an electron loss, while the yellow one indicates an electron gain.

Further, we have studied the electronic charge density difference $\Delta\rho(r)$ for a good understanding of the interaction nature of all the five gas molecules on the b-As surface. Practically, $\Delta\rho(r)$ illustrates the charge accumulation/depletion in the device by which we can calculate the total charge transfer. **Figure 3a-e** displays the computed charge transfer, which is computed by equation 2 for the adsorption each gas molecule (CO, CO$_2$, NO, NO$_2$, and NH$_3$) on the b-As surface. Our outcomes suggest that $E_{ads}$ depends upon the total charge transfer and redistribution between the adsorbed gas molecules and the b-As substrate. The charge transfer analysis is done by Bader charges linked with the adsorption of gas molecules on the b-As surface. The net charges on the adsorbed gases (in units of |e|) are tabulated in **Table 1**. For CO and $CO_2$ adsorption on b-As, one can see a small amount of charge (0.04 |e|) from the $CO/CO_2$ molecule to the b-As surface (**Figure 3a|b; Table 1**), resulting in a weak binding efficiency. When we have investigated the N-based gases, a significant charge transfer is found. Particularly for $NO_2$, which has the stronger $E_{ads}$ (**Figure 3d; Table 1**), up to 0.46 |e| charge is transferred from the b-As surface to the gas molecules. For the comparatively weaker $E_{ads}$ of $NO/NH_3$, the charge transfer is smaller than that of $NO_2$ but quiet more significant than those for $CO/CO_2$ [**Figure 3c|e; Table 1**]. Keeping other 2D monolayers such as graphene, phosphorene, $MoS_2$, and borophene in viewpoint our outcomes display reasonable $E_{ads}$ and appropriate charge transfer demonstrating better sensitivity for the latter. Our finding also delivers a road for electric field control of gas molecule adsorption, as revealed in $MoS_2$.[74]

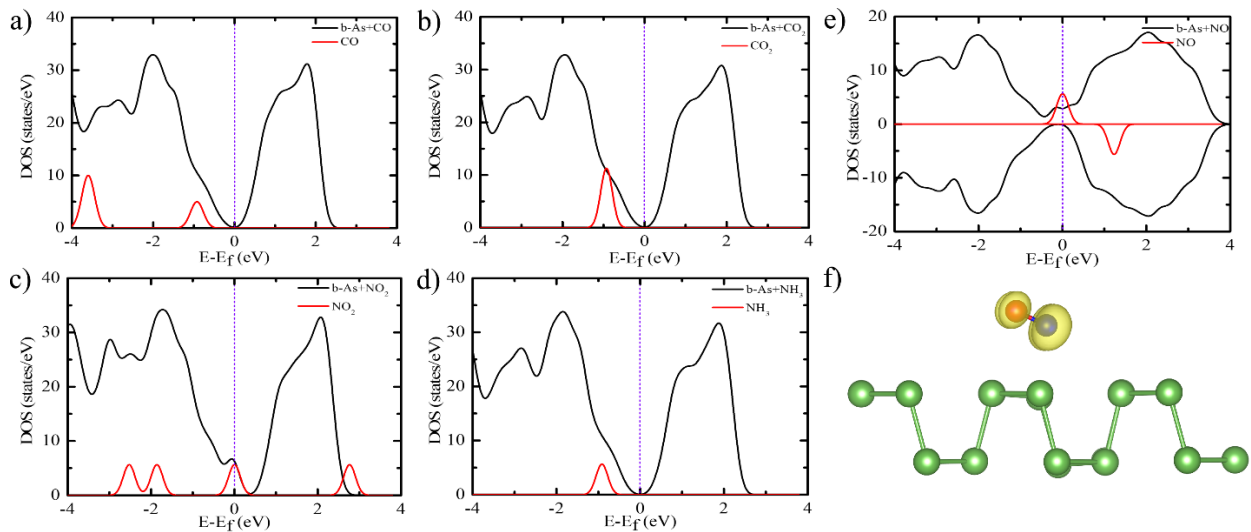

**Figure 4**. (a-d) TDOS of b-As with CO, CO$_2$, NO$_2$ and NH$_3$ gas adsorption (black lines) and the PDOS from the gas molecules (red lines). (e) Spin-polarized DOS of b-As with NO gas adsorption (black lines) and the PDOS from the NO gas (red lines) (f) The spatial-spin density distribution on NO gas. The Fermi level is aligned to zero (blue dashed lines).

Next, we have examined the effect of gas molecules adsorption on the electronic density of states (DOS) of the b-As surface. **Figure 4** displays the total DOS of the b-As+gas system along with the projected DOS on all five gas molecules. We find that the DOS of b-As is not considerably influenced due to the adsorption of CO, CO$_2$, and NH$_3$ gases, which is reliable with their smaller adsorption strength and total charge transfer. The gas molecules (NO and NO$_2$) are paramagnetic, and the adsorption of NO and NO$_2$ leads to a significant change in the DOS. **Figure 4e** shows the spin-polarized DOS figure of NO gas adsorption on the b-As surface. **Figure 4f** shows the spin density distribution of NO gas molecule adsorbed on the b-As surface. Different PDOS for up/down spin channels in NO case is observed. One can see peak broadening at the Fermi-level on PDOS for NO molecule due to the spin-up electrons. As shown in Figure 4f the spin-polarized electrons are located on the NO gas molecules. We have found the total magnetic moment is around 1 $\mu_B$ for NO molecule on the b-As. This magnetization occurs due to the existence of unpaired electrons in NO molecule. In the case of NO$_2$ molecules on b-As, we have computed both total DOS and PDOS of molecules (**Figure 4d**) and spin-polarized DOS (**Figure S1**; **SI**). It is noted that the NO$_2$ molecule adsorption on b-As manifest the highest adsorption energy leads to a magnetic moment of 0.03 $\mu_B$. From **Figure S1**, we have found similar spin-polarized DOS for up and down spin channels in NO case on the substrate. Hence from the foregoing discussion, it is appropriate to investigate more on the possibility of b-As based nano-scale devices for gas sensor applications. As our $E_{ads}$ values and significant charge transfer are shows that b-As based nano-scale device would fulfil our two prerequisites. In precise, the significant charger transfer is a sign that the interaction between the gas molecule and device is sufficient to yield noticeable changes in the electronic transport properties. Thus, to understand the enactment of b-As as a gas detector, the NEGF+DFT method is used to examine the electronic transport properties.

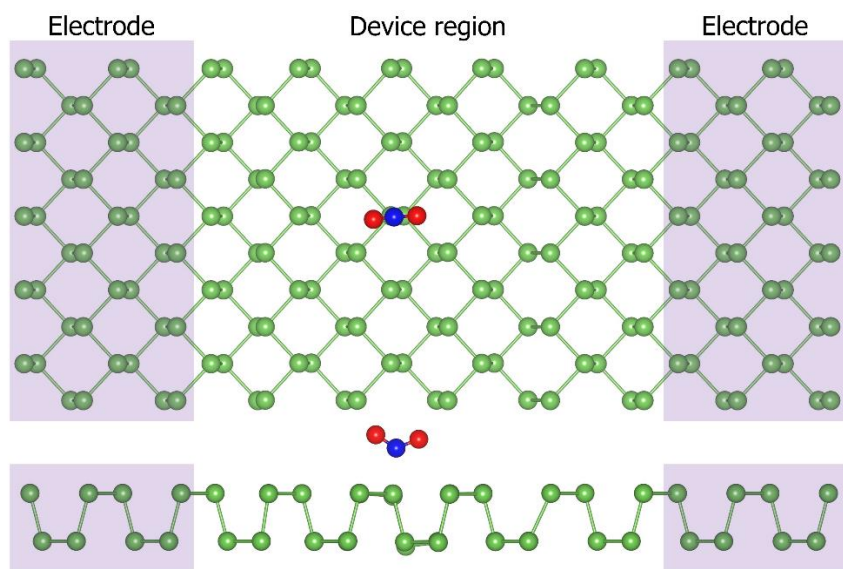

**Figure 5**. Schematic of the proposed nano-scale device set up (top and side views) displaying the semi-infinite electrode (left and right) regions and the device region.

Next, we explore the electronic transmission function and conductance sensitivity of b-As before and after the gas molecules adsorption. The transport setup is shown in **Figure 5**, where shaded areas depicts the electrode (left and right) and a device region where all the five gas molecules are adsorbed. Considering the fact that b-As has anisotropic transport behaviour, we have considered the armchair configuration for the device along the transport direction (z is the transport direction) as the electrical transmission of b-As is reported higher in the armchair direction than in the zigzag direction.[55] **Figure 6** shows the transmission spectrum of the pristine b-As and b-As+gas devices. These outcomes reveal that one can detect the gas molecules on their transmission signals around the Fermi-level. A noticeable effect of gas molecule adsorption in the transmission spectrum can be perceived. Further, from the experimental point of view, we have calculated the conductance changes of the device induced by the target gas molecule adsorption. The sensitivity (S%) is defined as; $S(\%) = |\frac{G_0-G}{G_0}|$, where $G_0$ is the conductance of the pristine b-As, and G is the conductance of the b-As+gas device. **Figure 6b|c** shows the conductance sensitivity at -1.275 eV, which can be experimentally measured by using a gate voltage of ±1.275 V. In principle, one may show the sensitivity for full energy spectra. The observed sensitivity values for negative (positive) gate voltage are 66.86%, 21.72%, 59.34%, 79.52%, and 75.04% (48.85%, 49.71%, 41.32%, 29.63%, and 1.54%) for CO, $CO_2$, NO, $NO_2$, and $NH_3$, respectively. We have found that CO, $CO_2$, NO, $NO_2$, and $NH_3$ can be detected at negative (positive) energies (i.e., gate voltages). At the negative gate voltage, the high sensitivity value for $NO_2$, leading to sensitivity over 79%. At the same gate voltage, the device is less than 4% sensitive in detecting $NH_3$. At the positive gate voltage, the proposed device displays the same detection capability for detecting CO and $CO_2$. The lowest sensitivity is noted for $NH_3$ at positive gate voltage. Note that the findings in **Figure 6** are also given as a function of the relevant $E_{ads}$. Consequently, it is not the influence of the strength of the interaction, rather the different type of gas molecules that controls the sensitivity. Overall, these outcomes indicate that b-As based nano-scale device possesses high sensitive resolution to detect each gas molecules electrically at a specific gate voltage. Lastly, regarding selectivity, we note that the adsorption energy values of all five gas molecules are sufficient enough to influence the resistance time together with electronic transport, therefore possibly leading to worthy selectivity as well.

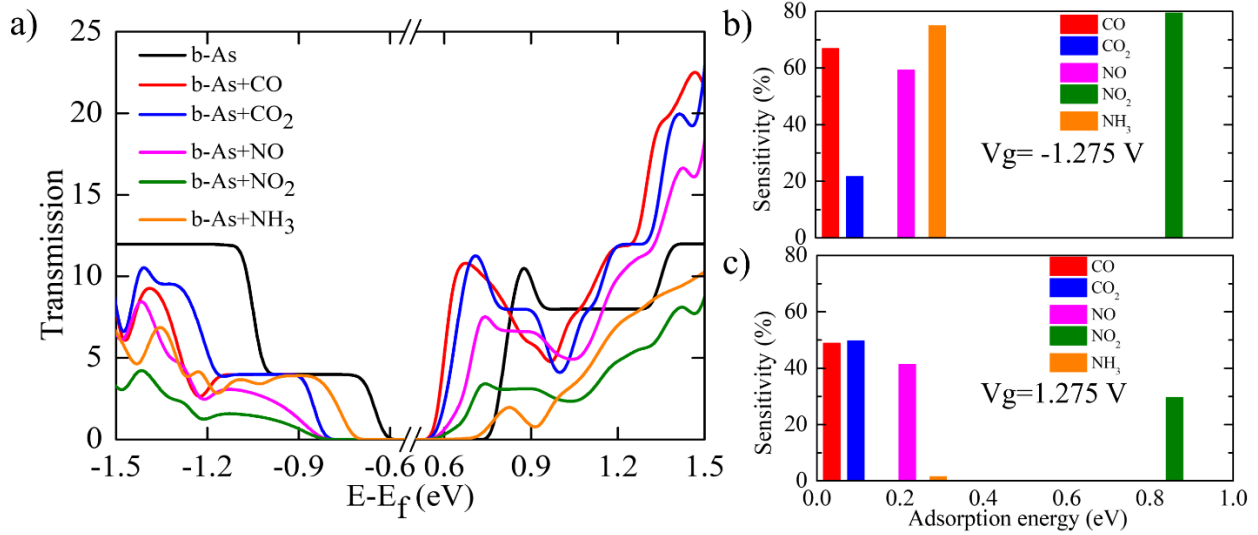

**Figure 6**. (a) The zero-bias transmission function in the vicinity of the Fermi level for the b-As (black line) and b-As+gas systems (colored lines); (b/c) sensitivity histograms for two specific energies (−1.275 and +1.275 eV).

## 4. Conclusions

In conclusion, we have demonstrated the gas sensing applications of recently reported 2D puckered black arsenic. Using first-principle calculations in combination with DFT and NEGF formalism, we have studied the structural, electronic, adsorption configuration, adsorption energy, charge density redistribution, and electronic transport properties of monolayer b-As with the adsorption of five gas molecules (CO, $CO_2$, NO, $NO_2$, and $NH_3$). Our findings show that the adsorption strength (energy) is optimum for all five gas molecules. The nitrogen (N) based gas molecules usually have the largest adsorption strength values than those for CO and $CO_2$ adsorption, demonstrating that b-As based nano-scale device is more sensitive to the NO, $NO_2$, and $NH_3$. This behavior leads to a good sensitive change in the electronic structure and charge transfer prompted by the adsorption of all the five gases. We find that the adsorption strength depends upon the charge transfer amount and redistribution between the gas molecules and the b-As substrate. After that, we have performed the electronic transport calculations to investigate the gas sensing capability if the b-As nano-scale device in terms of sensitivity and selectivity. Based on our outcomes, it has revealed that b-As based nano-scale device possesses high sensitive resolution to detect all the five (CO, $CO_2$, NO, $NO_2$, and $NH_3$) gas molecules electrically at a specific gate voltage. Thus, based on all these findings, such as sensitivity and selectivity to all the five gas molecules adsorption make black arsenic a promising material as an optimum gas sensor nano-scale device.

## 5. Associated Contents

*Supporting Information

## 6. Conflicts of interest

There are no conflicts of interest to declare.

## 7. Acknowledgments

We thank IIT Indore for the lab and computing facilities. This work is supported by DST-SERB, (Project Number: EMR/2015/002057) New Delhi and CSIR [Grant number: 01(2886)/17/EMR (II)] and (Project Number: CRG/2018/001131) and SPARC/2018-2019/P116/SL. R. L. K. and M. K. J. thanks MHRD for research fellowships. We would like to thank Dr. Vivekanand Shukla for fruitful discussion throughout this work.

## 8. ORCID

Rameshwar L. Kumawat: 0000-0002-2210-3428

Biswarup Pathak: 0000-0002-9972-9947